\documentclass{emulateapj}

\newcommand{\myemail}{srazzaqu@gmu.edu}

\newcommand{\be}{\begin{equation}}
\newcommand{\ee}{\end{equation}}
\newcommand{\ba}{\begin{eqnarray}}
\newcommand{\ea}{\end{eqnarray}}

\def\veps{\varepsilon}

\def\g{\gamma}

\def\p{\prime}

\shorttitle{UHECR and jet powers of TeV blazars}
\shortauthors{Razzaque, Dermer \& Finke}

\begin{document}

\title{Lower limits on ultrahigh-energy cosmic ray and jet powers of
TeV blazars}

\author{Soebur Razzaque\altaffilmark{1,2}, Charles
D.~Dermer\altaffilmark{3}, Justin D.~Finke\altaffilmark{3}}

\altaffiltext{1}{School of Physics, Astronomy and Computational
Sciences, George Mason University, Fairfax, Virginia 22030; \myemail}

\altaffiltext{2}{Present address: Space Science Division, U.S.~Naval
Research Laboratory, Washington, DC 20375}

\altaffiltext{3}{Space Science Division, U.S.~Naval Research
Laboratory, Washington, DC 20375}

\begin{abstract} 
Lower limits on the power emitted in ultrahigh-energy cosmic rays
(UHECRs), which are assumed to be protons with energy $\gtrsim
10^{17}$--$10^{20}$~eV, are derived for TeV blazars with the
assumption that the observed TeV $\gamma$ rays are generated due to
interactions of these protons with cosmic microwave photons.  The
limits depend on the spectrum of the injected UHECR protons.  While
for a $-2.2$ injection spectrum, the lower limits on the powers
emitted in UHECRs by 1ES~0229+200, 1ES~1101-232 and 1ES~0347-121 are
lower than their respective synchrotron luminosities ($\sim
10^{46}$~erg~s$^{-1}$); in the case of 1ES~1426+428 it exceeds the
corresponding synchrotron luminosity by up to an order of magnitude.
The proposed Auger North Observatory should be able to detect $4\times
10^{19}$~eV cosmic ray protons from 1ES~1426+428 within a few years of
operation and test the TeV $\g$-ray production model by UHECR energy
losses while propagating along the line-of-sight, or constrain
the intergalactic magnetic field to be larger than $\sim 10^{-16}$~G
in case of no detection.  The lower limits on the apparent-isotropic
jet power from accelerated $10^{10}$--$10^{20}$~eV proton spectra in
the blazar jet is of the order of the Eddington luminosity of a $10^9\
M_{\odot}$ black hole for a cosmic-ray injection spectrum $-2.2$ or
harder for all blazars considered except for 1ES~1426+428. In the case of
the latter the apparent-isotropic jet power exceeds the Eddington
luminosity by an order of magnitude.  For an injection spectrum softer
than $-2.2$, as is required to fit the observed cosmic-ray data above
$\sim 10^{17}$--$10^{18}$~eV, the Eddington luminosity is exceeded by
the lower limits on the jet power for all blazars considered.
\end{abstract}


\section{Introduction}

The origin and composition of the highest-energy cosmic rays (CRs) are
still unknown~\citep[see, e.g.,][]{gaiser10}.  They are generally
thought to be produced in extragalactic sources, due to a lack of
powerful Galactic sources capable of producing them and due to their
near-isotropic arrival direction distribution on large
scales~\citep{auger09}.  The Galactic supernova remnants are leading
candidates for producing CRs below $\sim 10^{17}$~eV~\citep[see,
  e.g.,][]{pzs10}.  Above this energy, denoted as ultrahigh energy
(UHE), observed CRs may originate in active galactic nuclei (AGNs)
with blazars being favored~\citep{rb93,Berezinsky06,drfa09,pmm09}
and/or in gamma-ray bursts~\citep[GRBs;][]{wax95,viet95,wda04,rdf10}
along with their low-luminosity counterparts~\citep{minn06,wrmd07}.

In the context of the AGN/blazar hypothesis, lack of a significant
correlation between the known position of the sources and the arrival
directions of the UHECRs~\citep{auger09} is generally explained by
deflections of the charged primaries in the Galactic and intergalactic
magnetic field (IGMF).  The value of the IGMF is not known but limits
exist, $\lesssim 10^{-6}$~G from cosmological observations~\citep[see,
  e.g.,][]{bfs97,kmk09}, and $\gtrsim 10^{-15}$--$10^{-16}$~G assuming
TeV blazars are persistent sources~\citep{nv10,tavecchio10b} while
$\gtrsim 10^{-17}$--$10^{-18}$~G assuming TeV blazars are
variable~\citep{dkot11,dcrfcl11,tvn11}.  UHECRs propagating in voids,
with very small value of the IGMF, may avoid significant deflection
and reach us directly from their sources, if nearby.  For sources
beyond $\sim 100$~Mpc, the highest-energy primary CRs cannot reach us
since they lose energy by interacting with the low-energy photons of
the Cosmic Microwave Background (CMB) and the Extragalactic Background
Light (EBL), a phenomenon known as the GZK
effect~\citep{greisen66,zk66}.  The secondary particles however, can
form an electromagnetic cascade and cascade radiation, mostly
Compton-scattered CMB photons~\citep[see,
  e.g.,][]{dzgmw02,rmz04,iit08,tavecchio10b,dcrfcl11}, can reach us as
if emitted from the source, if the IGMF value is sufficiently small.

Indeed it has been proposed recently that TeV $\g$-ray emission
detetcted from distant blazars and showing no significant flux
variation is secondary cascade emission by UHECR
protons~\citep{ek10,essey10,essey11}.  This proposition opens a
possibility to estimate the UHECR power and subsequently the jet power
of the TeV blazars, since protons are expected to dominate the energy
budget in such a scenario.  In an alternate scenario, non-variation of
TeV flux may arise from cascade emission in EBL and CMB from $\gtrsim
10$~TeV photons, which could be distinguished from proton-induced
cascade emission at very high energies~\citep{mdtm}. These models,
however, cannot explain rapid variability, often on hours time scale
and over broad frequency range, observed in the bulk population of
blazars and thought to be related to the size scale of the emission
region in the jet.  The broadband radio to $\g$ ray spectral energy
distribution of TeV blazars consists of two peaks: A low-energy
peak typically in the UV/X-ray and a high-energy peak at $~\sim
100$~GeV~\citep[see, e.g.,][]{gp94,TeVreview,tavecchio10a}.  While the
low-energy peak is generally modeled as synchrotron radiation, the
high-energy peak is often modeled as inverse Compton (IC) emission by
the sychrotron-radiating electrons~\citep[see,
  e.g.,][]{costamante02,boettcher07} or by hadronic
emission~\citep{muecke03}.  Apart from apparent non-variation of the
flux at $\gtrsim$~TeV in a number of high-redshift blazars, models of
high-energy $\g$-ray emission from the jet often requires very hard
intrinsic spectrum in order to avoid rapid flux reduction via $\g\g$
pair production with EBL photons~\citep[see,
  e.g.,][]{Fraceschini_EBL09,Finke_EBL10}.  Emission from UHECR
cascade in the EBL-CMB can potentially evade these
issues~\citep{essey10,essey11}.

In this paper we derive lower limits on the UHECR and jet powers of
the TeV blazars by assuming that the total energy loss rate,
integrated over energies $\gtrsim 10^{17}$~eV, by UHECR protons in the
electromagnetic channels ($p\gamma_{\rm CMB} \to e^+ e^-$;
$p\gamma_{\rm CMB}\to \pi^0 \to \g\g, ~\to \pi^\pm \to e^\pm$) and
after propagating to a redshift $z_\g$, when the universe becomes
transparent to TeV $\g$ rays, is equal to the bolometric TeV $\g$-ray
luminosity calculated from the observed spectra and $z_\g$.  For the
EBL models, not constrained by the {\em Fermi} Large Area Telescope
(LAT) data~\citep{fermi_EBL}, the $\g\g$ opacity $\tau_{\g\g} \approx
1$ at $z_{\g} \approx 0.1$.  For $z<z_\g$ cascade emission is not
reprocessed efficiently and is mostly emitted at very-high energies,
far above the $\sim 0.1$~TeV threshold of the air Cherenkov telescopes
but could potentially be detected by the proposed HiSCORE $\gamma$-ray
detector~\citep{HiSCORE} which would be sensitive in the PeV rage.

Our limits are based on the assumption that UHECR protons do not
suffer significant energy losses in the blazar jet and escape along
the direction within the jet opening angle.  As mentioned earlier, TeV
$\g$ rays that are produced in the jet from UHECR interactions, suffer
attenuation in the EBL.  Because of this reduction of the $\g$-ray
source flux, the total required UHECR power is greater if the observed
flux is a combination of the source and cascade fluxes than if solely
coming from the cascade.  We do not include deflections of the primary
proton and cascade $e^+ e^-$ in the IGMF, which reduce the flux of the
secondary radiation and increase the required UHECR and jet
powers~\citep[see, e.g.,][]{rmz04,dcrfcl11}.  Moreover, large $\gtrsim
10^{-9}$~G magnetic field is expected in the filaments in the galaxy
clusters~\citep{rkcd08} which can also reduce the cascade radiation
flux when encountered by the primary proton and cascade $e^+ e^-$,
thus further increasing the required UHECR and jet powers.  Our
limits, though model-dependent, are conservative in this sense.  They
will be weaker if other processes, e.g.\ leptonic emission from the
jet and/or from the cascade in EBL-CMB, contribute significantly to
the observed TeV data.  Based on our estimated lower limits on the
UHECR fluxes from the TeV $\g$-ray blazars, we also calculate the
detection rate of the primary UHECR protons, which travel along the
line-of-sight from their sources, by the Pierre Auger
Observatory.\footnote{\url{http://www.auger.org/}}

We calculate the energy losses by UHECR protons in the CMB in Sec.~2,
and estimate the apparent-isotropic luminosity of TeV blazars from
spectra measured by air Cherenkov telescopes in Sec.~3.  We derive our
limits on the UHECR and jet powers in Sec.~4, and calculate UHECR
event rate based on the limits in Sec.~5.  We discuss results and
conclude our investigation in Sec.~6.

\section{Cosmic ray spectra and energy losses}

We calculate the power-law source spectrum (total number of particles
generated isotropically from the source per unit energy and time
intervals) with an index $-\kappa$ and between the generation energies
$E_{1g}$--$E_{2g}$, given an apparent-isotropic bolometric
luminosity (power) of cosmic rays $L_{\rm CR}$, as
\be
q(E_{g}) \equiv \frac{dN}{dE_{g}dt} = 
\frac{L_{\rm CR}(\kappa -2)E_{1g}^{\kappa-2}}
{1-(E_{1g}/E_{2g})^{\kappa -2}} \,E_{g}^{-\kappa} ~;~ 
\kappa \ne 2.
\label{source_spectum}
\ee
The observed energy of a CR on the Earth, however, is degraded due to
adiabatic or redshift (due to expansion of the universe) loss and due
to collisional losses with the CMB and/or EBL photons.  The
relationship between the generated CR energy $E_g(z)$ at redshift $z$
and the observed energy $E$ is~\citep{Berezinsky06}
\ba
E &=& E_g(z) 
- \int_{0}^{z} dz^\p ~\frac{E_g (z^\p)}{1+z^\p} 
\nonumber \\ && 
- \int_{0}^{z} dz^\p ~\frac{1+z^\p}{H(z^\p)} 
~b_0 [(1+z^\p) E_g(z^\p)], 
\label{past_present}
\ea
where $E_g (z^\p) = E_g(z)(1+z^\p)/(1+z)$, $H(z^\p)$ is the comoving
Hubble parameter and $b_0(E) = -dE/dt$ is the energy loss rate of an
UHECR at the present epoch ($z=0$).  UHECR protons lose their energy
by interacting dominantly with the CMB photons, creating $e^+e^-$
pairs and $\pi$'s.  These energy-loss rates and their evolution with
redshift have been calculated by many authors, and they are well
understood in the $10^{17}$--$10^{20}$~eV energy range~\citep[see,
e.g.,][]{stanev00,Berezinsky06,dm10}.
 
The flux of cosmic rays on the Earth from a source at redshift $z$ is
then
\ba
J(E) &=& \frac{q(E_{g})}{4\pi d_c^2 (1+z)} \frac{dE_{g}}{dE} 
\nonumber \\
&\approx & \frac{L_{\rm CR}(\kappa -2)E_{1}^{\kappa-2}}
{4\pi d_c^2 (1+z)^{3-\kappa}} E^{-\kappa}  
\left( \frac{E_{g}}{E} \right)^{-\kappa}
\frac{dE_{g}}{dE} ~;~ \kappa\ne 2,~~~
\label{CR_flux}
\ea
where the second line follows with an approximation $E_{1g} \ll
E_{2g}$ in Equation~(\ref{source_spectum}) and when the lowest-energy CRs
lose energy due to redshift only, i.e.\ $E_{1g} = E_1 (1+z)$.  The
Jacobian factor $dE_{g}/dE$ follows from Equation~(\ref{past_present}) and
is given by~\citet{Berezinsky06}.  The comoving distance $d_c$, which
is related to the proper distance $d_p$ and the luminosity distance
$d_L$ as $d_L = (1+z) d_c = (1+z)^2 d_p$, is defined as~\citep[see,
e.g.,][]{hogg99}
\be
d_c = \frac{c}{H_0} \int_0^z \frac{dz^\p}
{\sqrt{\Omega_m (1+z^\p)^3 + \Omega_\Lambda}} \,.
\label{co_distance}
\ee
We assume the standard cosmological parameters: $\Omega_m = 0.27$,
$\Omega_\Lambda = 0.73$ and a Hubble constant $H_0 =
71$~km~s$^{-1}$~Mpc$^{-1}$.  

Figure~1 shows the UHECR proton spectra as would be observed from a
source at various redshifts but with the same apparent-isotropic
luminosity $L_p = 10^{45}$~erg~s$^{-1}$ for $10^{17}$~eV~$\le E_g \le
10^{20}$~eV and for the spectral index of the generated protons to be
$\kappa = 2.2$.  Note that the sharp cutoff at the highest energies
$\gtrsim 4\times 10^{19}$~eV is due to $\pi$ losses (the GZK effect)
and the dip below $\sim 4\times 10^{19}$~eV but above $\sim
10^{18}$~eV is due to $e^+ e^-$ pair
losses~\citep{Berezinsky05}. These spectral features shift to lower
energy for high-redshift sources.  Adiabatic losses dominate at the
lowest energy in this plot.

\begin{figure}
\begin{centering}
\includegraphics[width=3.2in]{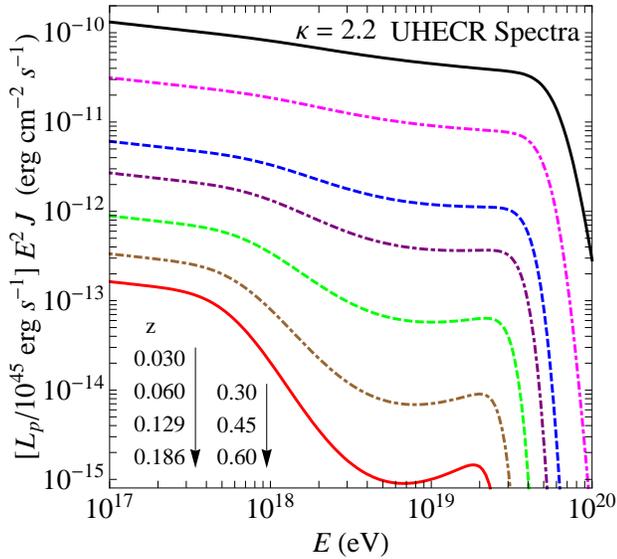}
\caption{
Expected UHECR proton spectra on the Earth from a source with
increasing redshift from the top to bottom according to
Equation~(\ref{CR_flux}).  We assumed an apparent-isotropic UHECR source
luminosity to be $10^{45}$~erg~s$^{-1}$ with a power-law spectrum of
index $\kappa = 2.2$ in the generated energy range
$10^{17}$--$10^{20}$~eV.  Adiabatic (redshift), $e^+e^-$ pair and
$\pi$ creation losses have been taken into account in the numerical
calculation.  }
\end{centering}
\label{fig:UHECR_spectra}
\end{figure}

The total bolometric power lost, due to creating $e^+e^-$ pairs and
$\pi$'s, by the UHECR protons, generated at a redshift $z$ and
measured at another redshift $z_0$, can be calculated from
Equation~(\ref{source_spectum}) and Equation~(\ref{past_present}) considering
only collisional losses and by integrating over the generation energy
in the range $E_{1g}$--$E_{2g}$.  The total power lost in
electromagnetic (EM) channels is smaller when considering energy
losses to secondary charged particles and photons only, which can
initiate electromagnetic cascades by interacting with the CMB and EBL
photons.  Below $E_g\sim 4\times 10^{19}$~eV, energy losses by UHECRs
are mostly in the EM channels.  Above $E_g\sim 4\times 10^{19}$~eV,
roughly $1/3$ of the bolometric power lost, with $\sim 50\%$
inelasticity, goes into producing neutrinos. Subtracting the neutrino
losses we calculate the ratio of the power lost in EM channels, due to
propagation, to the generated UHECR power $L_{\rm CR}$ as
\ba
f_{\rm CR} &=&
\frac{(\kappa -2) E_{1g}^{\kappa -2}}
{1-(E_{1g}/E_{2g})^{\kappa -2}} \frac{c}{H_0}
\frac{(1+z_0)^2}{(1+z)^2}
\int_{E_{1g}}^{E_{2g}} dE_g\, E_g^{-\kappa} \nonumber \\
&& \times \int_{z_0 = z_\g}^{z} dz^\p 
\frac{(1+z^\p)\, b_{0,\rm EM}[(1+z^\p) E_g(z^\p)]}
{\sqrt{\Omega_m (1+z^\p)^3 + \Omega_\Lambda}}.
\label{loss_fraction}
\ea
Here the lower limit of the redshift integration corresponds to $z_\g$
at which $\tau_{\g\g}\approx 1$, since efficient reprocessing of the
EM energy losses by UHECRs to observed TeV $\g$ rays can take place
only above this redshift.

Figure~2 shows the fraction $f_{\rm CR}$ as a function of the redshift
and for different values of the spectral index $\kappa$.  The top and
bottom panels correspond to the $10^{17}$--$4\times 10^{19}$~eV and
$10^{17}$--$10^{20}$~eV energy ranges of the generated spectra,
respectively.  The dotteded curves in each panel correspond to the
present redshift ($z_0 = 0$) in Equation~(\ref{loss_fraction}).  The
solid curves correspond to the redshift $z_0 = z_\g = 0.1$ at which
$\g\g\to e^+ e^-$ pair production opacity $\tau_{\g\g} = 1$ at TeV,
according to the EBL models by~\citet{Finke_EBL10} and
by~\citet{Fraceschini_EBL09}, and those photons can propagate to us
without much attenuation~\citep{fs70}. Figure~3 shows $f_{\rm CR}$ as
in Fig.~2 but above $10^{18}$~eV for comparison.  Note that the loss
fraction increases in this case because a smaller part of the UHECR
spectrum is considered here, where collisional losses are the most
important.  As can be seen, the loss fraction crucially depends on the
index $\kappa$ of the generated CR spectrum and is smaller for softer
spectra.

\begin{figure}
\begin{centering}
\includegraphics[width=3.6in]{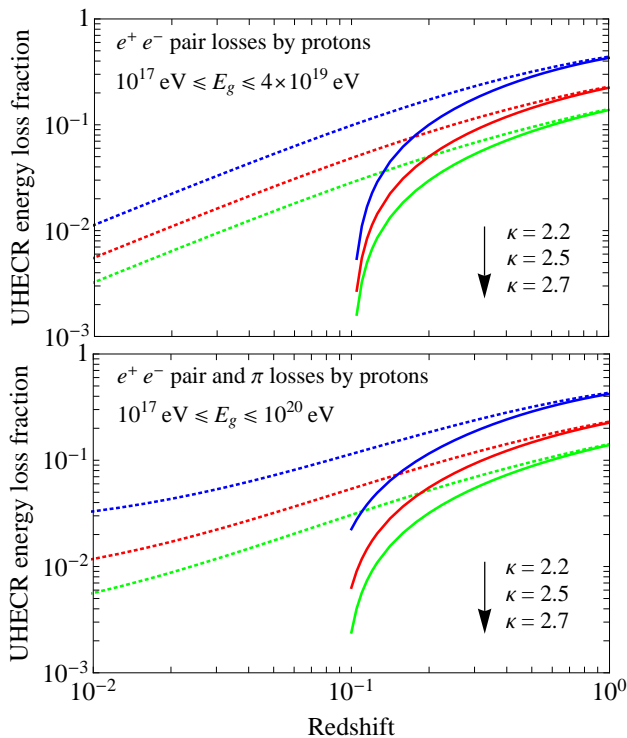}
\caption{
Total bolometric power loss fraction by UHECR protons, generated at
various redshifts and with different spectral index $\kappa$,
propagting to $z=0$ (dotted curves) and to $z=0.1$ (solid curves). The
universe becomes optically thin ($\tau_{\g\g}\lesssim 1$) to TeV
photons due to $e^+e^-$ pair creation with the EBL photons at
$z\approx 0.1$~\citep[see, e.g., the EBL models
by][]{Finke_EBL10,Fraceschini_EBL09} {\em Top panel ---} Energy losses
by $e^+e^-$ pair creation, which dominates below $\approx 4\times
10^{19}$~eV, with the CMB photons.  {\em Bottom panel ---} Energy
losses by $e^+e^-$ pair and $\pi$ creation, which dominates above
$\approx 4\times 10^{19}$~eV, with the CMB photons.  The lowest UHECR
energy is assumed to be $10^{17}$~eV.}
\end{centering}
\label{fig:ECR_loss}
\end{figure}

\begin{figure}
\begin{centering}
\includegraphics[width=3.6in]{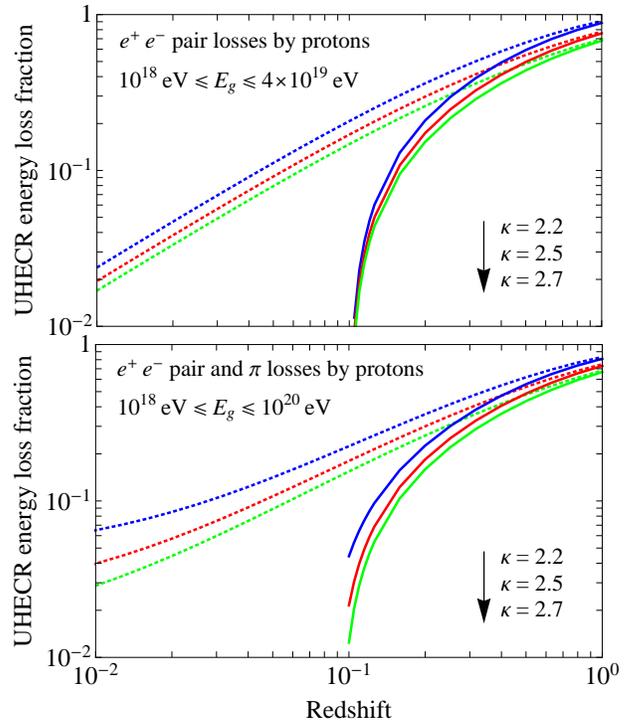}
\caption{
The same as in Fig.~2 but the lowest UHECR energy is assumed to be
$10^{18}$~eV.}
\end{centering}
\label{fig:ECR_loss2}
\end{figure}

The Lorentz factor of the $e^+e^-$ pairs created from a $10^{18}$~eV
proton is roughly $\g_e \approx 10^9$.  The Compton-scattered CMB
photon energy by the pairs is $(4/3)\g_e^2 \varepsilon_{\rm CMB}
\approx 900\,(1+z)(\g_e/10^9)^2$~TeV.  Subsequent absorption of these
photons in the EBL, creating $e^+e^-$ pairs, and Compton scattering of
CMB photons give rise to an EM cascade.  A similar scenario applies
for $\pi^0\to \g\g$ and $\pi^\pm \to e^\pm$ channels.  Compton photons
of a given energy from the cascade can reach the Earth once their
opacity falls below unity.  Thus substantial reprocessing of UHECR
power lost in EM channels to observed TeV $\g$ rays can only take
place at $z_\g\gtrsim 0.1$.  We estimate the total EM power loss by
UHECRs from TeV $\g$-ray data in the next section.

\section{TeV Blazars and Gamma Ray Luminosity}

We assume that the TeV $\g$ rays detected from blazars at $z\gtrsim
z_\g \approx 0.1$ are reprocessed cascade emission from UHE protons
with energy $10^{17} \le E_g \le 10^{20}$~eV, that are generated at
the blazars.  TeV emission should not be rapidly variable in this case
since cascade emission takes place over a large size scale.  There can
be some variability at an energy much below 1~TeV, due to
contamination by the source photons.  Currently there are four such
TeV blazars with adequate data, namely 1ES~1426+428~\citep{hess1426},
1ES~0229+200~\citep{hess0229,veritas0229},
1ES~1101-232~\citep{hess1101} and 1ES~0347-121~\citep{hess0347}.
These are all BL Lac objects.

Figure~4 shows the very high-energy $\g$-ray data collected from these
blazars.  None of these sources except 1ES~1101-232 has been
convincingly detected\footnote{A tentative detection of 1ES~0229+200
  has been reported by~\citet{okd11} and by~\citet{fermi_1FGL}}, which
is sensitive in the 0.1--300~GeV range. However data at much lower
frequencies, Optical-UV and X-ray, are available.  Emission at these
frequencies is widely believed to be synchrotron radiation from
non-thermal electrons in the blazar jet. The bolometric synchrotron
luminosity $L_{\rm syn}$, calculated as ten times the specific
luminosity $\nu L_{\nu}$ at the peak of the synchrotron emission as
given in~\citet{tavecchio10a}, is also listed in in Table~1 for each
of the four blazars.

\begin{figure*}
\vskip 0.1in
\begin{centering}
\includegraphics[width=4.3in,angle=270]{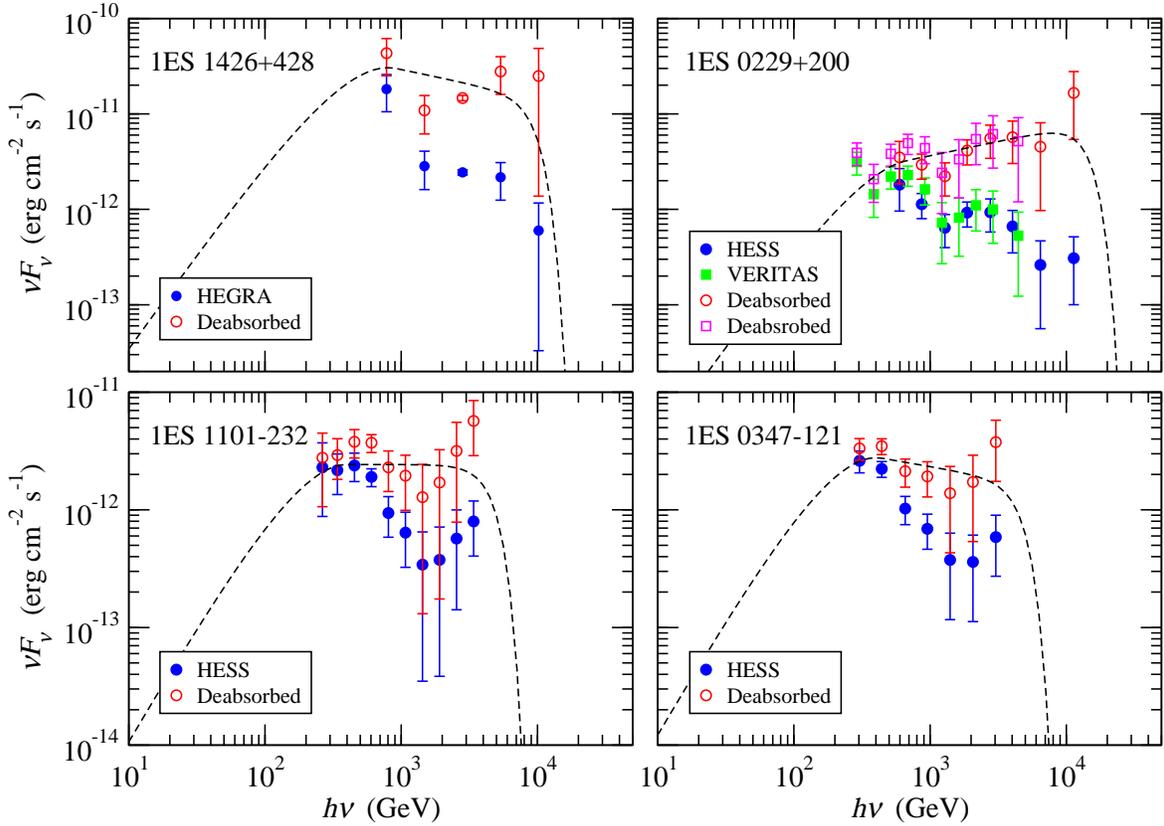}
\caption{
TeV $\g$-ray data (filled points) from the HESS and VERITAS
observations on selected blazars at $z\gtrsim 0.1$.  Fits (dashed
lines) to the $\g$-ray data from these blazars are performed after
deabsorption (empty points) at $z=0.1$ with the EBL model
by~\citet{Finke_EBL10}.  The fit function consists of two smoothly-
joined power laws with a super-exponential, $\exp(-\nu^4)$, cutoff at
high energies as in Equation~(\ref{fit}).  The low-energy slope of the
power law is kept fixed so that $\nu F_\nu \propto \nu^2$, analogous
to the hardest IC spectrum. The high-energy index $\beta$ is listed in
Table 1.}
\end{centering}
\label{fig:blazar_spectra}
\end{figure*}

\begin{table}[b]
\caption{Blazar and TeV $\g$-ray model parameters}
\begin{tabular}{ccccc}
\hline \hline
Blazar         & $z$   & $L_{\rm syn}$ & $L_{z_\g = 0.1}$ & $\beta$ \cr
               &       &  (erg/s)      & (erg/s)          &         \cr
\hline
1ES~$1426+428$ & 0.129 & $6\times 10^{45}$ & $2.1\times 10^{45}$ & $-2.3$ \cr
1ES~$0229+200$ & 0.139 & $2\times 10^{46}$ & $4.6\times 10^{44}$ & $-1.7$ \cr
1ES~$1101-232$ & 0.186 & $3\times 10^{46}$ & $2.1\times 10^{44}$ & $-2.0$ \cr
1ES~$0347-121$ & 0.188 & $2\times 10^{46}$ & $2.0\times 10^{44}$ & $-2.2$ \cr
\hline
\end{tabular}
\end{table}

Figure~4 also shows the deabsorbed $\g$-ray data points using
the~\citet{Finke_EBL10} EBL model at $z_\g = 0.1$.  We fit the
deabsorbed $\g$-ray spectra (black dotted lines) by two
smoothly-joined power laws, similar to a Band spectrum~\citep{band93},
with a super-exponential cutoff at the high energies as below
\ba
\frac{dN}{d\veps dA dt} = C \cases{\veps_{\rm GeV}^\alpha
~\exp[\frac{(2+\alpha)\veps}{\veps_{\rm pk}}] 
~;~ \veps < \frac{(\alpha -\beta)\veps_{\rm pk, GeV}}{2+\alpha} \cr
\veps_{\rm GeV}^\beta 
~[\frac{(\alpha-\beta)\veps_{\rm pk, GeV}}
{2+\alpha}]^{\alpha-\beta} \cr
~~\times e^{\beta-\alpha} e^{-(\veps/\veps_c)^4}
~;~ \veps \ge \frac{(\alpha -\beta)\veps_{\rm pk, GeV}}{2+\alpha} .
}
\label{fit}
\ea
We keep the index $\alpha$ of the low-energy power law fixed such that
$\nu F_\nu \propto \nu^2$ ($\alpha = 0$), the hardest Compton spectrum
expected, to be conservative.  The break energy $\veps_{\rm pk}$, at
which the $\nu F_\nu$ spectrum turns over, the high-energy power-law
index $\beta$ and the cutoff energy $\veps_{c}$ were allowed to vary.

Finally we calculate a conservative lower limit on the
apparent-isotropic TeV $\g$-ray luminosity from our fitted spectra
(see Fig.~4, dashed lines) to the deabsorbed data points at $z_\g =
0.1$ as
\be
L_{z_\g = 0.1} = 4\pi d_L^2 (z_\g) \int d\nu ~F_\nu
\label{gamma_lum}
\ee
in the 10~GeV--20~TeV range.  This luminosity and the high-energy
photon index $\beta$ in Equation~(\ref{fit}) along with the synchrotron
luminosity for individual blazars are listed in Table 1.

\section{Limits on UHECR and jet powers}

We calculate the lower limits on the apparent-isotropic UHECR power
of the TeV blazars from $\g$-ray luminosity calculated in
Equation~(\ref{gamma_lum}) as
\be
L_{\rm UHECR} > L_{z_\g = 0.1} /f_{\rm CR}
\label{Luhecr}
\ee
Note that for high EBL models~\citep[see, e.g.,][]{stecker06}, with
$z_\g < 0.1$, $L_{z_\g}$ will be lower.  A lower $z_\g$ also gives a
higher $f_{\rm CR}$ or more efficient conversion of UHECR energy to EM
energy.  Thus our limits will be weaker for high EBL models, a result
consistent with what~\citet{essey11} found by modeling TeV $\g$-ray
data.

The top panel of Fig.~5 shows the lower limits on the
apparent-isotropic UHECR power, in the $10^{17}$--$10^{20}$~eV range,
of individual blazars that we have derived from the TeV $\g$-ray
luminosity $L_{z_\g=0.1}$ listed in Table~1 and Equation~(\ref{Luhecr}).
The limits are in the range $10^{45}$--$10^{47}$~erg/s and are
consistent with the required UHECR source luminosity from detailed
modeling of TeV $\g$-ray data~\citep{essey10}.

The bottom panel of Fig.~5 shows the lower limits on the blazar jet
power that we have derived from the TeV $\g$-ray data, and are plotted
against the synchrotron luminosities of the corresponding blazars.  To
calculate these lower limits we have extrapolated the proton spectrum
down to a minimum energy of $10$~GeV.  Thus the bulk Lorentz factor of
the relativistic TeV blazar jets need to be $\sim 10$, which is
consistent with measured values.  The empty symbols (as in the top
panel) correspond to the limits assuming the same power-law index
$\kappa$ as above $10^{17}$~eV and as is expected in a Fermi
shock-acceleration scenario.  The filled symbols correspond to the
limits assuming $\kappa=2.2$ below $10^{17}$~eV and $\kappa=2.5$ or
2.7 above $10^{17}$~eV.  Such broken power-law spectra may arise in
more complicated scenarios, e.g.\ from cooling of UHECR protons in the
jet. In case the minimum proton energy is $100$~GeV, our limits will
be weaker by factors 1.6, 3.2 and 5.0 for $\kappa = 2.2$, 2.5 and 2.7
respectively, in case of single power-law spectrum.  Also shown
(dotted line) is the Eddington luminosity $L_{\rm Edd} = 1.3\times
10^{47}~(M_{\rm bh}/10^9~M_\odot)$~erg/s of a $10^9~M_\odot$
blackhole, which is the typical black-hole mass thought to power the
blazar jets.

In case the TeV blazars produce UHECRs above $10^{18}$~eV rather than
above $10^{17}$~eV, our limits on the UHECR power becomes lower (see
Fig.~6 top panel) by a factor of $\sim 2$--5, depending on the
injection spectral index $\kappa$.  This is an expected result since
the blazars need to power a smaller part of the UHECR spectrum in this
case.  The lower limits on the jet power (Fig.~6 bottom panel) in case
of broken power-law spectra (filled symbols) are also reduce by a
factor $\sim 2$--3 while shifting the break energy from $10^{17}$~eV
to $10^{18}$~eV, for similar reasons as for the limits on the UHECR
power.  Our limits on the jet power for single power-law spectra
(empty symbols in Fig.~6 bottom panel) from $10^{10}$~eV to
$10^{20}$~eV, however, remains unchanged.

\begin{figure*}
\begin{centering}
\includegraphics[trim=0.in 0.in 0.1in 0.in,clip=true,width=5.2in]{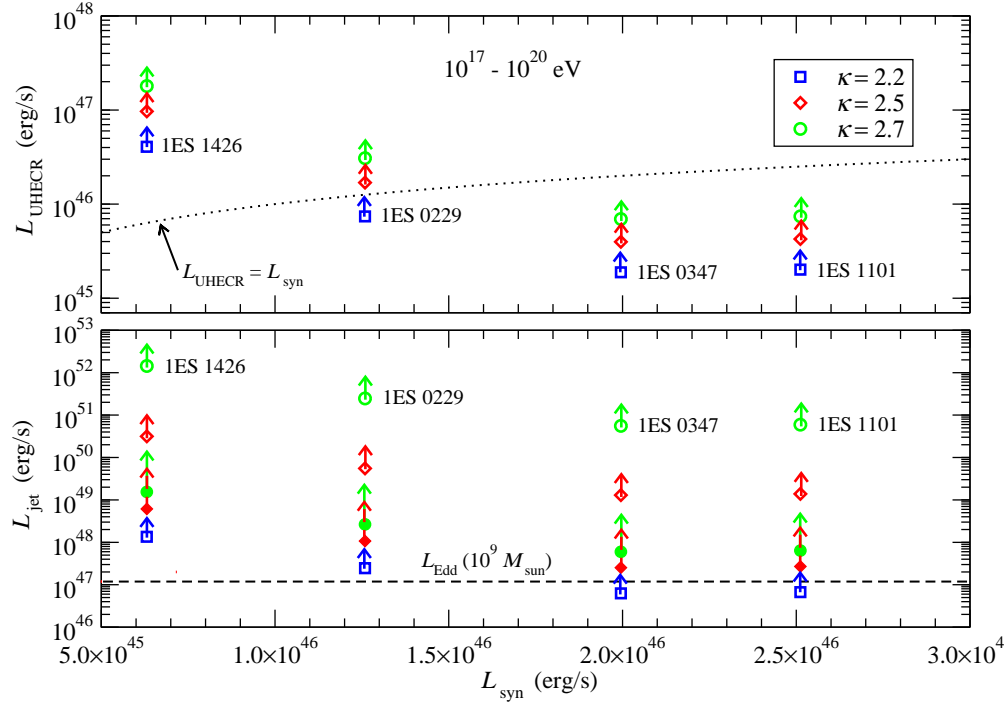}
\caption{ 
{\em Top panel---} Lower limits on the apparent-isotropic UHECR power,
in the $10^{17}$--$10^{20}$~eV range and assuming different injection
spectral index $\kappa$, from TeV blazars derived from the TeV
$\g$-ray luminosity $L_{z_\g = 0.1}$ in Table 1.  The limits on the
UHECR power are plotted against the synchrotron power (mostly in X
rays) of the respective blazars, and the two powers are equal at the
dotted line.  {\em Bottom panel---} Lower limits on the
apparent-isotropic jet power derived from TeV $\g$-ray data and
plotted against their respective synchrotron luminosities.  The jet
powers are calculated from the CR spectra in the
$10^{10}$--$10^{20}$~eV range by extrapolating the spectrum below
$10^{17}$~eV, with single power law index $\kappa$ as above
$10^{17}$~eV (empty symbols), and with broken power law of fixed index
$\kappa = 2.2$ below $10^{17}$~eV but varying $\kappa$ above
$10^{17}$~eV (filled symbols). Also shown (dashed line) is the
Eddington luminosity of a $10^9~M_\odot$ blackhole.}
\end{centering}
\label{fig:uhecr_limit}
\end{figure*}

\begin{figure*}
\begin{centering}
\includegraphics[trim=0.in 0.in 0.1in 0.in,clip=true,width=5.2in]{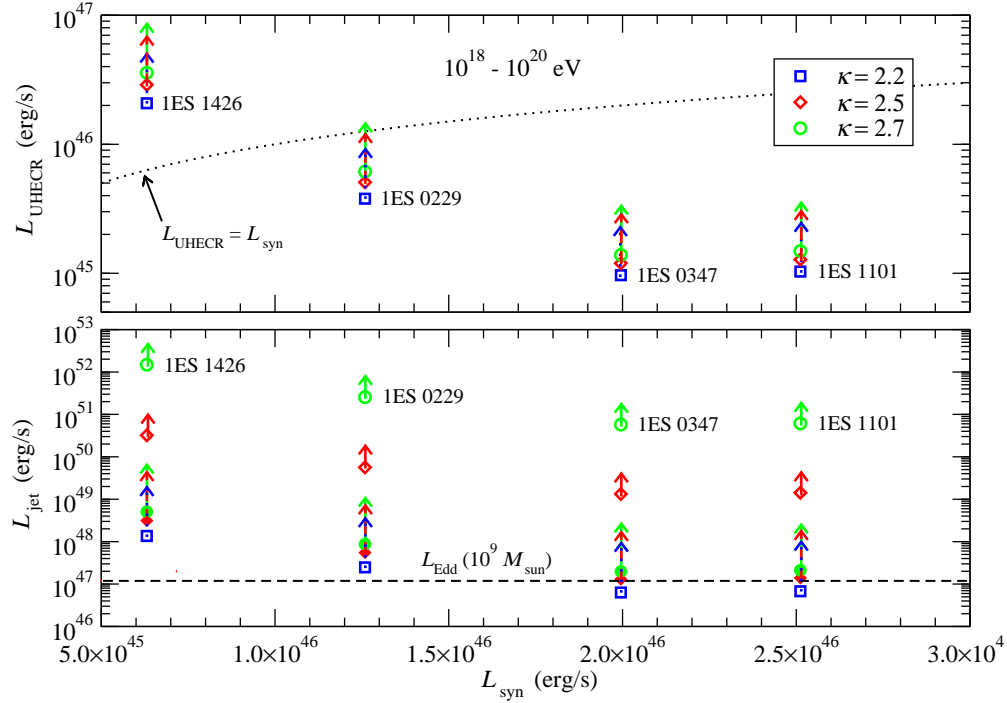}
\caption{ 
The same as in Fig.~5
but assuming that UHECRs are
generated in the $10^{18}$--$10^{20}$~eV range.}
\end{centering}
\label{fig:uhecr_limit2}
\end{figure*}

The limits that we have derived are based on the assumption that
primary UHECRs and the cascade emission are concentrated within the
opening angle of the blazar jet, $\theta_{\rm jet} \approx 0.1$.  For
$10^{17}$~eV CRs and a characteristic propagation distance of $d_{\rm
p} \approx 500$~Mpc with $l_{\rm coc} \approx 1$~Mpc coherence length
scale of IGMF, this implies that the mean IGMF strength should be
\ba
B_{\rm IGMF, UHECR} &\lesssim & 10^{-12} 
\left( \frac{\theta_{\rm jet}}{0.1} \right)
\left( \frac{E}{10^{17}~{\rm eV}} \right)
\nonumber \\ && \times
\left( \frac{d_p}{500~{\rm Mpc}} \right)^{-1/2}
\left( \frac{l_{\rm coh}}{{\rm Mpc}} \right)^{-1/2}
~{\rm G},~~~
\label{igmf_uhecr}
\ea
for the deflection angle to be smaller than $\theta_{\rm jet}$.  A
stronger constraint on the IGMF, however, comes from the requirement
that the cascade electrons of Lorentz factor $\g_e \approx 10^6$ that
upscatter CMB photons to TeV $\g$ rays do not deviate out of
$\theta_{\rm jet}$.  The cooling length scale of electrons in nearby
universe is $\lambda_{\rm IC} = 0.71~(\g_e/10^6)^{-1}$~Mpc, and to
avoid any deflection larger than $\theta_{\rm jet}$ one requires
\ba
B_{\rm IGMF, TeV} \lesssim 10^{-16} 
\left( \frac{\theta_{\rm jet}}{0.1} \right)
\left( \frac{\veps_\g}{{\rm TeV}} \right)^2
~{\rm G}.
\label{igmf_tev}
\ea

To avoid significant deflection in the Mpc scale structured region of
galaxy clusters with $\sim 10^{-9}$~G field~\citep{rkcd08,mdtm}, the
coherence length scale should be $\lesssim 10$~kpc from
Equation~(\ref{igmf_uhecr}).  For much larger magnetic field and/or
coherence length the UHECRs become isotropic and their luminosity
decreases by a beaming factor $f_b = (1-\cos\theta_{\rm jet})^{-1}
\approx 200$ for $\theta_{\rm jet} = 0.1$.  As a result our lower
limits on UHECR power will become stronger in such a case.

Propagation of UHECRs from the TeV blazars nearly along our
line-of-sight, in low IGMF environment, brings up naturally the
question of their detectability by the UHECR detectors on the Earth.
While the highest-energy CRs are absorbed during propagation in the
CMB, since the blazars are beyond the GZK radius, sufficiently
energetic CRs can still reach us (see Fig.~1) and we discuss the
prospect of their detection in the next section.

\section{Detectability of UHECRs from TeV Blazars}

The location of the Pierre Auger Observatory~\citep{auger04} at
latitude $35.2^\circ$~S makes it insensitive to the nearest and most
powerful UHECR candidate source 1ES~1426+428 with a Declination
$\delta = +42.67^\circ$.  The proposed Auger North
detector\footnote{\url{http://www.augernorth.org/}} at latitude
$39^\circ$ N, however, will be sensitive to 1ES 1426+428.  For a
detector similar to the Auger South, the instrument exposure becomes
independent of energy at $E\gtrsim 3\times 10^{18}$~eV and depends
only on the zenith angle $\theta_z$~\citep{auger08}. For Auger South
$\theta_z \le 60^\circ$. The relative exposure at a point source in
the sky, such as 1ES~1426+428, compared to the largest exposure on the
sky~\citep{sommers01} is $\omega (\delta) = 0.62$ for the Auger North.
The actual exposure on the source can be found from multiplying
$\omega (\delta)$ by the total integrated exposure ${\cal E}$
(km$^2$~yr~sr) over the detector's field-of-view and dividing by the
solid angle of the detector $\Omega = \sqrt{3}\pi$~sr, for $\theta_z
\le 60^\circ$~\citep[see also][]{ch08}.  Thus the total number of
UHECR events expected from 1ES~1426+428, for $\kappa = 2.2$ source
spectrum, in the proposed Auger North detector is
\ba
N_{\rm evt} &=& \frac{{\cal E}\omega (\delta)}{\Omega} 
\int_{E_{\rm th}}^{E_{\rm mx}} J(E)~dE
\nonumber \\ &\approx &
3 \left( \frac{L_p}{10^{46}~{\rm erg/s}} \right)
\left( \frac{\cal E}{9,000~{\rm km^2~yr~sr}} \right),
\ea
above a threshold energy $E_{\rm th} = 4\times 10^{19}$~eV and below
the maximum energy $E_{\rm mx} = 10^{20}$~eV.  Thus 1ES~1426+428
should be clearly visible, if an UHECR source as modeled here, with
the same exposure used in the first data release by the Auger South
Observatory~\citep{auger08}.

The arriving UHECRs from 1ES~1426+428 are expected to be deflected by
the Galactic magnetic field~\citep[see, e.g.,][]{drfa09} by an angle
\ba
\theta_{\rm Gal} &=& 1.5^\circ 
\left( \frac{E}{4\times 10^{19}~{\rm eV}} \right)^{-1}
\left( \frac{B_{\rm Gal}}{10^{-6}~{\rm G}} \right)
\nonumber \\ && \times
\left( \frac{h_{\rm disc}}{{\rm kpc}} \right)
\left( \frac{\sin b}{\sin 64.9^\circ} \right)^{-1},
\ea
where $h_{\rm disc} \sim 1$~kpc is the height of the Galactic disc and
$b=64.9^\circ$ is the galacic latitude of 1ES~1426+428.  The
deflection angle while propagating in the IGMF, with $B_{\rm IGMF,
TeV} \lesssim 10^{-16}$~G, is negligible.  Thus UHECRs from
1ES~1426+428 should be clustered around the source direction for a
Northern site detector similar to Auger South with $\sim 1^\circ$ or
better resolution above $\sim 10^{19}$~eV~\citep{auger08}.

It is interesting to note that the position of 1ES~1426+428 is within
$\sim 5^\circ$--10$^\circ$ of three clustered (triplet) events
detected by the AGASA experiment during its operation with energies
$4.97\times 10^{19}$~eV, $4.98\times 10^{19}$~eV and $5.27\times
10^{19}$~eV~\citep{agasa99,agasa00}.  The average angular resolution
of AGASA however is $1.8^\circ$, smaller than the angular separation
between the triplet and 1ES~1426+428.  Downgrading the event energies
by $25\%$~\citep{agasa03} does not increase the deflection angle
$\theta_{\rm Gal}$ significantly.  If, however, $B_{\rm Gal} \sim
4\times 10^{-6}$~G in the Galactic disc, $\theta_{\rm Gal}$ can be
large enough to encompass the nearest AGASA event from the triplet.
Although the exposure is lower compared to its peak exposure in other
directions in the sky, HiRes experiment has not detected any event
above $4\times 10^{19}$~eV from the direction of
1ES~1426+428~\citep{hires_aniso_10}.

The detection prospect for UHECRs with $\kappa = 2.2$ from the other 3
blazars within the field-of-view of the Auger South is much smaller,
$N_{\rm evt} = 0.04$,~0.03 and 0.02 respectively for 1ES~0229+200
($\delta = +20.27^\circ$), 1ES~1101-232 ($\delta = -23.50^\circ$) and
1ES~0347-121 ($\delta = -11.98^\circ$); for $L_p = 10^{45}$~erg/s and
${\cal E}= 9,000$~km$^2$~yr~sr.

\section{Discussion and Conclusions}

The lower limits on the UHECR power $L_{\rm UHECR}$, in the
$10^{17}$--$10^{20}$~eV range, that we have derived are lower than the
synchrotron luminosity $L_{\rm syn}$ of the sources for $\kappa =
2.2$, except for 1ES~1426+428 where $L_{\rm UHECR}/L_{\rm syn} \sim
6$,~15, and 30 for $\kappa = 2.2$,~2.5, and 2.7 respectively. While
none of these sources have been detected by the {\em Fermi}
LAT~\citep{fermi_1LAC}, $L_{\rm UHECR}$ is a factor $\sim 1$--60 times
the median $\g$-ray luminosity of $\sim 3\times 10^{45}$~erg/s, in the
LAT range, from the 118 BL Lacs detected by {\em Fermi} with known
redshift~\citep{fermi_1LAC}.  The ratio $L_{\rm UHECR}/L_{\rm syn}$
gets smaller in the case TeV blazars generate cosmic rays in the
$10^{18}$--$10^{20}$~eV range only.

The lower limits on the jet power that we have calculated, exceed
$L_{\rm syn}$ and the median BL Lac $\g$-ray luminosity in the {\em
Fermi} LAT range for all the blazars we considered.  In the most
extreme case of 1ES~1426+428 the ratio $L_{\rm jet}/L_{\rm syn} \sim
2\times 10^2,~5\times 10^4,~2\times 10^6$ for the single power-law
spectrum with $\kappa = 2.2$,~2.5,~2.7 respectively.  Even for $\kappa
= 2.2$, the limiting Eddington luminosity is violated in this case by
an order of magnitude.  For other blazars, $L_{\rm jet} \lesssim
L_{\rm Edd}$ for $\kappa = 2.2$ and single power-law spectrum.
However $L_{\rm jet} \gtrsim L_{\rm Edd}$ for $\kappa > 2.2$, both for
the single power-law and broken power-law spectra.

If the observed TeV $\g$ rays from the blazars that we have considered
are due to UHECR acceleration in these sources and due to their energy
losses while propagating along the line-of-sight, the source spectrum
of UHECRs needs to be $\propto E^{-2.2}$ or harder in order for the
model to be energetically viable.  If the synchrotron luminosity is a
good indicator of the contribution by the relativistic electrons to
the jet power, then the power in baryons with Lorentz factor $\gamma_p
\ge 1$ in the relativistic jet needs to be larger by a factor of 200
or more in the case of 1ES~1426+428 with $\kappa = 2.2$.  The actual
jet power will be even higher when including non-relativistic baryons
in the jet.  The proposed Auger North Observatory should be able to
detect CRs above $4\times 10^{19}$~eV from 1ES~1426+428 within a few
years of operation and test the model of TeV $\g$ ray origin from
UHECRs as proposed by~\citet{essey10,essey11} and as discussed here.
Detection of GZK neutrinos from 1ES~1426+428 by the currently
operating IceCube neutrino observatory at the South Pole and by its
proposed extensions could be possible.  Non-detection of UHECRs from
1ES~1426+428 can imply a large IGMF as generally inferred from a lack
of significant correlation between the sources and UHECR arrival
directions, and/or favor leptonic production mechanism of TeV $\g$
rays.

To fit the observed UHECR spectrum as measured by the Auger and HiRes
arrays, with protons, an injection spectrum $\propto E^{-2.7}$ above
$10^{18}$~eV is required to explain the dip and cutoff as due to
$e^+e^-$ pair production and $\pi$ production,
respectively~\citep{Berezinsky08}.  A broken power law $\propto
E^{-2.7}$ above and $\propto E^{-2.2}$ below a break-energy $\sim
10^{17}$--$10^{18}$~eV can help reduce the energy crisis in the AGN
interpretation of the observed UHECRs.  This scenario also makes the
energy requirement for the UHECR interpretation of TeV $\g$ rays from
blazars less severe.  To match the observed spectrum, the total energy
injection rate in UHECRs above $10^{17}$~eV in the local universe
needs to be $\sim 10^{45}$~erg~Mpc$^{-3}$~yr$^{-1}$~\citep[see
e.g.][]{vdg03,Berezinsky06}.  The energy injection rates above
$10^{17}$~eV from the 4 TeV blazars that we considered are $(\gtrsim
2, \gtrsim 6, \gtrsim 11)\times
10^{45}f_b^{-1}$~erg~Mpc$^{-3}$~yr$^{-1}$, respectively, for $\kappa =
(-2.2, -2.5, -2.7)$ from our lower limits on UHECR power.  The energy
injection rates are calculated from the UHECR powers divided by the
comoving volumes corresponding to the redshifts of each
blazar~\citep{dr10}.  If the structured region with high magnetic
field near a TeV blazar makes $\gtrsim 10^{17}$~eV cosmic rays
isotropic, our lower limits on UHECR power will be higher by a factor
$\sim f_b$ and the corresponding energy injection rate could be
sufficient to explain the observed UHECR spectrum above $\sim
10^{17}$~eV.  There will be many TeV $\g$-ray sources without the
UV/X-ray synchrotron counterparts from the misaligned blazars in this
scenario, which could be searched for with existing and future
telescopes.

The jet power exceeds the Eddington luminosity in almost all cases we
considered and will do so more severely if the UHECRs become isotropic
in the scenario discussed above.  It is possible that extragalactic
sources with hard injection spectrum contribute to the observed UHECRs
far above $\sim 10^{17}$~eV, as explored in the GRB model with $\kappa
\sim 2.2$ above the ankle at $\sim 4\times 10^{18}$~eV and with
observed star-formation rate~\citep{da06}.  However, in such models an
additional component of UHECRs in the intermediate $\sim
10^{17}$--$4\times 10^{18}$~eV energy range is required.  In principle
they can originate from low-luminosity GRBs/hypernovae of either
extragalactic~\citep{wrmd07} or Galactic~\citep{wda04,bkmw08,calvez10}
origin.  Whether nature conspires to generate such a high-energy break
in the injection spectrum or the observed UHECRs originate in multiple
source classes is a matter of ongoing investigations.  However, the
interpretation of TeV $\g$ rays from blazars generating UHECRs seems
less favorable in case the injection spectrum is similar to the one
required to explain the observed cosmic-ray spectrum above $\sim
10^{17}$--$~10^{18}$~eV.

\section*{Acknowledgments}

We thank M.~B{\"o}ttcher, C.~C.~Cheung, L.~Costamante, M.~Mostafa and
K.~Murase for helpful comments and discussion.  We also thank the
anonymous referee for a constructive report.  This work is supported
by grants from the NASA Fermi Cycle 3 guest investigator program and
by the Office of Naval Research.  Work of S.R. was funded while under
contract with the U.S.~Naval Research Laboratory.

\end{document}